\documentclass[preprint,10pt,1p,sort&compress]{elsarticle}
\pdfoutput=1
\usepackage{amssymb,amsmath}
\usepackage{latexsym,mathrsfs}
\usepackage{graphicx,subfigure}
\usepackage{epsfig,varioref}
\usepackage{xr-hyper,color,multirow}
\usepackage{tikz,array,wasysym}
\usepackage{xcolor,float}
\usepackage{diagbox}
\usepackage[utf8]{inputenc}

\journal{Phys. Dark Univ.}

\begin{document}

\begin{frontmatter}

\title{The galaxy power spectrum take on spatial curvature and cosmic concordance}

\author[kicc]{Sunny Vagnozzi\corref{cor1}}
\ead{sunny.vagnozzi@ast.cam.ac.uk}\cortext[cor1]{Corresponding author}

\author[durham,manchester]{Eleonora Di Valentino}
\ead{eleonora.di-valentino@durham.ac.uk}

\author[turin,ific]{Stefano Gariazzo}
\ead{gariazzo@to.infn.it}

\author[rome]{Alessandro Melchiorri}
\ead{alessandro.melchiorri@roma1.infn.it}

\author[ific]{Olga Mena}
\ead{omena@ific.uv.es}

\author[iap,jhu,oxford]{Joseph Silk}
\ead{joseph.silk@physics.ox.ac.uk}

\address[kicc]{Kavli Institute for Cosmology, University of Cambridge, Cambridge CB3 0HA, UK}
\address[durham]{Institute for Particle Physics Phenomenology, Durham University, Durham DH1 3LE, UK}
\address[manchester]{Jodrell Bank Center for Astrophysics, University of Manchester, Manchester M13 9PL, UK}
\address[turin]{Istituto Nazionale di Fisica Nucleare, Sezione di Torino, I-10125 Turin, Italy}
\address[ific]{Instituto de F\'{i}sica Corpuscular, University of Valencia-CSIC, 46980 Valencia, Spain}
\address[rome]{Department of Physics and INFN, University of Rome ``La Sapienza'', I-00185 Rome, Italy}
\address[iap]{Institut d'Astrophysique de Paris, CNRS/UPMC Universit\'{e} Paris 6, F-75014, Paris, France}
\address[jhu]{Department of Physics and Astronomy, The Johns Hopkins University, Baltimore, MD 21218, USA}
\address[oxford]{BIPAC, Department of Physics, University of Oxford, Keble Road, Oxford OX1 3RH, UK}

\begin{abstract}
\noindent The concordance of the $\Lambda$CDM cosmological model in light of current observations has been the subject of an intense debate in recent months. The 2018 \textit{Planck} Cosmic Microwave Background (CMB) temperature anisotropy power spectrum measurements appear at face value to favour a spatially closed Universe with curvature parameter $\Omega_K<0$. This preference disappears if Baryon Acoustic Oscillation (BAO) measurements are combined with \textit{Planck} data to break the geometrical degeneracy, although the reliability of this combination has been questioned due to the strong tension present between the two datasets when assuming a curved Universe. Here, we approach this issue from yet another point of view, using measurements of the full-shape (FS) galaxy power spectrum, $P(k)$, from the Baryon Oscillation Spectroscopic Survey DR12 CMASS sample. By combining \textit{Planck} data with FS measurements, we break the geometrical degeneracy and find $\Omega_K=0.0023 \pm 0.0028$. This constrains the Universe to be spatially flat to sub-percent precision, in excellent agreement with results obtained using BAO measurements. However, as with BAO, the overall increase in the best-fit $\chi^2$ suggests a similar level of tension between \textit{Planck} and $P(k)$ under the assumption of a curved Universe. While the debate on spatial curvature and the concordance between cosmological datasets remains open, our results provide new perspectives on the issue, highlighting the crucial role of FS measurements in the era of precision cosmology.
\end{abstract}


\end{frontmatter}

\section{Introduction}
\label{sec:intro}

What is the shape of the Universe? This one simple question has plagued scientists, philosophers, and humankind in general, since the dawn of time, and relates to two aspects. The first is the Universe's global geometry, as described by its topology, and the second is the observable Universe's local geometry. The latter can be characterized by measuring the so-called \textit{spatial curvature} of the Universe, quantifying how much the spatial geometry locally differs from that of flat space. Cosmological observations can constrain the spatial curvature of the Universe, and in particular the so-called curvature parameter $\Omega_K$, measuring the effective fractional contribution of spatial curvature to the energy budget today, with $\Omega_K=0$ corresponding to spatial flatness~\cite{Dodelson:2003ft}.

In a large-scale isotropic and homogeneous Universe, $\Omega_K$ plays a fundamental role, because it has both a crucial role in determining the evolution of the Universe, and a close connection to early-Universe physics. In fact, most models of inflation~\cite{Kazanas:1980tx,Starobinsky:1980te,Guth:1980zm,Sato:1981ds,Mukhanov:1981xt,Linde:1981mu,Albrecht:1982wi} predict a Universe which is extremely close to being spatially flat~\cite{Baumann:2009ds,Martin:2013tda}. A convincing measurement of $\vert \Omega_K \vert \gtrsim 10^{-2}$ could spell trouble for many inflationary models~\cite{Linde:2007fr}, including most models of eternal inflation~\cite{Kleban:2012ph,Guth:2012ww} (see however~\cite{Bull:2013fga} for a different point of view). The fact that significant efforts have been devoted towards constraining spatial curvature (hereafter simply ``curvature'') from cosmological observations and forecasting the achievable precision on such constraints from future observations (see e.g.~\cite{Ichikawa:2006qb,Gong:2006gs,Clarkson:2007bc,Mao:2008ug,Virey:2008nu,Vardanyan:2009ft,Barenboim:2009ug,Mortsell:2011yk,Carbone:2011bx,Li:2012vn,Dossett:2012kd,Farooq:2013dra,Bull:2014rha,Takada:2015mma,Chen:2016eyp,DiDio:2016ykq,Moresco:2016nqq,Leonard:2016evk,Yu:2016gmd,Verde:2016ccp,Rana:2016gha,Ooba:2017ukj,Cao:2017rfk,Ooba:2017npx,Witzemann:2017lhi,Yu:2017iju,Jimenez:2017tkk,Ooba:2017lng,Park:2017xbl,Denissenya:2018zcv,Park:2018bwy,Wei:2018cov,Aghanim:2018eyx,Park:2018fxx,Park:2018tgj,Abbott:2018xao,Bernal:2018myq,Xu:2018kbm,Li:2019qic,Eingorn:2019spk,Jesus:2019jvk,Park:2019emi,Handley:2019tkm,Bianchini:2019vxp,Wang:2019yob,DiValentino:2019qzk,Zhai:2019nad,Geng:2020upf,Kumar:2020ole,Efstathiou:2020wem,Heinesen:2020sre,DiValentino:2020hov,Gao:2020irn,Bose:2020cjb,Khadka:2020hvb,Nunes:2020uex,Liu:2020pfa,Chudaykin:2020ghx,Shirokov:2020dwl,Benisty:2020otr,Shimon:2020dvb,Cao:2021ldv} for an inevitably incomplete list) should therefore come as no surprise.

Ever since the 1990s, with Cosmic Microwave Background (CMB) measurements from BOOMERanG~\cite{Melchiorri:1999br, deBernardis:2000sbo} and MAXIMA~\cite{Balbi:2000tg}, evidence has been accumulating in favor of our Universe being flat to within $10\%$, with such an indication becoming progressively refined with more precise data from WMAP~\cite{Hinshaw:2012aka}. However, the latest measurements of CMB temperature and polarization anisotropies from the \textit{Planck} satellite legacy data release~\cite{Akrami:2018vks,Aghanim:2018eyx,Aghanim:2019ame} \textit{might} be challenging this view: at first glance, these measurements (P18 hereafter) appear to point towards a Universe which is spatially closed, with a 99\% probability region of $-0.095 \leq \Omega_K \leq -0.007$~\cite{Aghanim:2018eyx}. This indication for spatial curvature, however, is significantly reduced to about $1.7$ standard deviations when CMB lensing power spectrum measurements are included, and does not survive when P18 is complemented with Baryon Acoustic Oscillation (BAO) data~\cite{Aghanim:2018eyx}. The inclusion of distance moduli measurements from Type Ia Supernovae (SNeIa) also points towards a flat Universe. In this case, however, the assumption of dark energy being a cosmological constant is key: relaxing this assumption leads once more to the preference for a closed Universe from the P18+SNeIa combination, as discussed in~\cite{DiValentino:2020hov}.

There is no doubt that a confirmed genuine detection of spatial curvature from P18 (a dataset which is extremely mature from both the theoretical and observational points of view) would imply nothing short of a crisis for modern cosmology, not only because of its inconsistency with basic inflationary predictions,~\footnote{In general, it is easier to construct inflationary models leading to open Universes (see~\cite{Coleman:1980aw,Gott:1982zf,Ratra:1994ni,Ratra:1994vw,Ratra:1994dm,Bucher:1994gb,Linde:1995xm,Yamamoto:1995sw,Linde:2007fr,Kleban:2012ph,Guth:2012ww} for early work), for instance in models where the Universe arises from quantum tunnelling-induced false vacuum decay, whereas constructing inflationary models leading to closed Universes might require more fine-tuning~\cite{Ratra:1984yq,Hartle:1983ai,Linde:2003hc,Ratra:2017ezv}.} but perhaps more importantly because of the inconsistency with the independent BAO, SNeIa, and CMB lensing datasets. It is at this point worth recalling that the validity of the concordance $\Lambda$CDM model is already being challenged by mild-to-strong tensions across independent inferences of cosmological parameters. Notable among these are tensions between independent inferences of the Hubble constant $H_0$~\cite{Aghanim:2018eyx,Riess:2019cxk,Wong:2019kwg,Freedman:2019jwv,Pesce:2020xfe}, an ongoing crisis usually referred to as the ``Hubble tension''~\cite{Verde:2019ivm,DiValentino:2020zio}. The Hubble tension has fueled an ongoing discussion as to whether physics beyond $\Lambda$CDM is required to address this discrepancy. While a satisfying solution has yet to be found, several models of new physics have been proposed, with varying degree of plausibility and/or ability to address the tension, see e.g.~\cite{DiValentino:2016hlg,Bernal:2016gxb,Karwal:2016vyq,Kumar:2016zpg,Kumar:2017dnp,Zhao:2017urm,DiValentino:2017iww,Sola:2017znb,Buen-Abad:2017gxg,Yang:2017ccc,Khosravi:2017hfi,Benetti:2017juy,Mortsell:2018mfj,Vagnozzi:2018jhn,Nunes:2018xbm,Poulin:2018zxs,Kumar:2018yhh,Yang:2018euj,Banihashemi:2018oxo,DEramo:2018vss,Guo:2018ans,Graef:2018fzu,Banihashemi:2018has,Aylor:2018drw,Poulin:2018cxd,Kreisch:2019yzn,Raveri:2019mxg,Martinelli:2019dau,Kumar:2019wfs,Agrawal:2019lmo,Li:2019san,Yang:2019jwn,Keeley:2019esp,Lin:2019qug,Li:2019ypi,Gelmini:2019deq,Rossi:2019lgt,DiValentino:2019exe,Archidiacono:2019wdp,Kazantzidis:2019dvk,Desmond:2019ygn,Yang:2019nhz,Nesseris:2019fwr,Pan:2019gop,Vagnozzi:2019ezj,Visinelli:2019qqu,Cai:2019bdh,Pan:2019hac,Xiao:2019ccl,Knox:2019rjx,DiValentino:2019ffd,Smith:2019ihp,Benetti:2019lxu,Ghosh:2019tab,Sola:2019jek,Escudero:2019gvw,Yan:2019gbw,Arendse:2019hev,Banerjee:2019kgu,Yang:2019uog,DiValentino:2019jae,Niedermann:2019olb,Sakstein:2019fmf,Anchordoqui:2019amx,Hart:2019dxi,Frusciante:2019puu,Akarsu:2019hmw,Ye:2020btb,Krishnan:2020obg,Lucca:2020zjb,DAgostino:2020dhv,Hogg:2020rdp,Benevento:2020fev,Hill:2020osr,Desmond:2020wep,Gomez-Valent:2020mqn,Akarsu:2020yqa,Ballesteros:2020sik,Haridasu:2020xaa,Alestas:2020mvb,Jedamzik:2020krr,Braglia:2020iik,Chudaykin:2020acu,Jimenez:2020bgw,Ballardini:2020iws,Aljaf:2020eqh,Ivanov:2020mfr,DiValentino:2020naf,Elizalde:2020mfs,Gu:2020ozv,Keeley:2020rmo,Elizalde:2020pps,Yang:2020myd,Efstathiou:2020wxn,Benaoum:2020qsi,Calderon:2020hoc,Ye:2020oix,Vazquez:2020ani,Akarsu:2020vii,LinaresCedeno:2020uxx,Murgia:2020ryi,Choudhury:2020tka}. In any case, if confirmed, the possible ``curvature tension''~\cite{DiValentino:2020srs} which is the subject of this paper, in conjunction with the Hubble tension, could be the first significant challenge to the otherwise extremely successful $\Lambda$CDM model~\cite{DiValentino:2020hov}.

Given the high stakes, there has been no shortage of discussion in the literature as to the physical significance of these results, as for instance in~\cite{Handley:2019tkm,DiValentino:2019qzk,Efstathiou:2020wem}. Handley in~\cite{Handley:2019tkm} (H19 hereafter) and Di Valentino \textit{et al.} in~\cite{DiValentino:2019qzk} (dV19 hereafter) pointed out that \textit{Planck} observations favour a closed Universe at high significance, possibly implying a crisis for modern cosmology. A similar result was already present in the Planck parameters paper.~\footnote{See Pages~30 and~40 of Ref.~\cite{Aghanim:2018eyx}.} On the other hand, Efstathiou and Gratton in~\cite{Efstathiou:2020wem} (EG20 hereafter) argue that these results are partly a consequence of both the choice of \textit{Planck} likelihood used,~\footnote{While the Planck collaboration, H19, and dV19 use the \texttt{Plik} likelihood, EG20 uses a modified version of the \texttt{CamSpec} likelihood~\cite{Efstathiou:2019mdh}, referred to as \texttt{12.5HMcl}. This likelihood is argued to be statistically more powerful than the standard \texttt{CamSpec} likelihood, since it has access to a larger sky fraction in both temperature and polarization.} as well as an over-interpretation of the $\Omega_K$ posterior: the latter is highly sensitive to the choice of prior on $\Omega_K$, and EG20 argued that it is dangerous to interpret it as a probability distribution unless one can strongly justify the choice of prior, which most (if not all) works have usually taken to be uniform on $\Omega_K$.

Part of the current debate about the Universe's spatial curvature is, therefore, ultimately centered on the differences between the \texttt{Plik} and \texttt{CamSpec} likelihoods. There are a number of differences between the two, including the treatment of polarization data. For more detailed discussions, we encourage the reader to consult Refs.~\cite{Aghanim:2019ame,Efstathiou:2019mdh}, and in particular Sections~3.5.1 and~6.3 thereof respectively. Using the \texttt{12.5HMcl} \texttt{CamSpec} likelihood, EG20 find a 99\% probability region for the curvature parameter of $-0.083<\Omega_K<-0.001$, which EG20 argue could be consistent with a statistical fluctuation.

However, besides the choice of \textit{Planck} likelihood and treatment polarization data therein, another major bone of contention in this debate is the treatment of external data, \textit{i.e.}\ data other than P18's temperature and polarization anisotropy measurements. EG20 correctly argue that including BAO measurements leads to a strong preference for a flat Universe, regardless of the \textit{Planck} likelihood used. On the other hand, both H19 and dV19 argue that \textit{within the assumption of a non-flat Universe} such a dataset combination should be viewed with caution, due to the mutual disagreement between the datasets involved. Such a concern is basically stating the view that before two or more datasets can be safely combined, they should be consistent, \textit{i.e.}\ plausibly arise from the same realization of our Universe. Using the so-called suspiciousness statistic, H19 finds the tension between P18 temperature and polarization data, CMB lensing, and BAO within a non-flat Universe to be $2.5$-$3\sigma$, which agrees with the findings of dV19. On the other hand it was argued in~\cite{Vagnozzi:2020dfn} that the combination of \textit{Planck} data and cosmic chronometer measurements indicates that the Universe is spatially flat, while not incurring in the aforementioned tensions.

Moving to CMB data coming from experiments other than \textit{Planck}, it is also worth remarking that, after combining their latest DR4 results with WMAP data and adopting the usual flat prior on $\Omega_K$, the Atacama Cosmology Telescope (ACT) collaboration finds a 68\% probability region of $-0.011 \leq \Omega_K \leq 0.013$, remarkably consistent with $\Omega_K=0$~\cite{Aiola:2020azj}. Similar results are obtained with ACT data alone, whereas combining ACT with \textit{Planck} leads to a 68\% probability region of $-0.028<\Omega_K<-0.005$, with the corresponding 95\% probability region instead encompassing $\Omega_K=0$. However, the combination of \textit{Planck} and ACT should be viewed with some caution, due to tensions at the $2.5\sigma$ level between the two, discussed both in the main ACT paper~\cite{Aiola:2020azj} and in~\cite{Handley:2020hdp}. In any case, the ACT results confirm that, by measuring the lensing of the CMB to very high accuracy, it is possible to break the $\Omega_K$-$\Omega_m$ geometrical degeneracy, a finding which is consistent with the \textit{Planck} simulations performed in~\cite{DiValentino:2019qzk}. We will return to the geometrical degeneracy, and its implications for spatial curvature, later in the paper.

This paper does not seek to take sides between the two different views in the debate. In fact, we wish to note that both arguments have their merits. On the other hand, we also note that in the ongoing curvature debate, several cosmological observations have been consulted: from the CMB lensing power spectrum reconstructed from the temperature 4-point function, to BAO and SNeIa distance measurements, to local distance ladder measurements of $H_0$, calibrated both with Cepheid variables and with the Tip of the Red Giant Branch. Full-shape galaxy power spectrum measurements have also been consulted (although to a lesser extent than BAO measurements): examples are by the BOSS collaboration making use of configuration space and Fourier space clustering wedges in~\cite{Sanchez:2016sas,Grieb:2016uuo}, by the DES collaboration through a joint $3\times2$pt analysis of galaxy clustering and weak lensing data in~\cite{Abbott:2018xao}, by the eBOSS collaboration in~\cite{Alam:2020sor}, and from an independent re-analysis in~\cite{Chudaykin:2020ghx}. In this work, we shall also listen to what the full-shape galaxy power spectrum measurements have got to tell us regarding spatial curvature, and more generally regarding the concordance between different cosmological datasets when moving beyond a spatially flat Universe. In particular, we shall present a re-analysis of the BOSS DR12 full-shape galaxy power spectrum independent from all the previously mentioned analyses, based on the earlier works of some of us in~\cite{Giusarma:2016phn,Vagnozzi:2017ovm,Giusarma:2018jei}, focusing on spatial curvature and the concordance with \textit{Planck} data within a non-flat Universe.

Studies of the clustering of tracers of the large-scale structure (LSS), such as galaxies, quasars, or the Lyman-$\alpha$ forest, have usually focused on the BAO signature contained within the configuration space correlation function $\xi(r)$ and the Fourier space power spectrum $P(k)$. However, it is possible to explore the full cosmological information content of the shape of $\xi(r)$ or $P(k)$ rather than just focusing on the BAO signature therein. In this work, focusing on the galaxy power spectrum $P(k)$, we will perform this type of full-shape (FS hereafter) analysis. Being strongly correlated, in principle FS and BAO measurements from a given LSS survey should not be used at the same time, unless one can correctly model their cross-covariance (as done in e.g.~\cite{Philcox:2020vvt}).

FS measurements can contain \textit{in principle} more information than BAO measurements: whether or not this is the case depends on several variables such as redshift of the sample, range of wavenumber modes analysed, reconstruction efficiency, and so on. At the same time, their theoretical modelling, especially in the mildly non-linear regime, is more challenging. This is perhaps the reason why the analysis of FS measurements has mostly only been performed within the context of large collaborations, with only a few exceptions (see e.g.~\cite{Escudero:2015yka,Cuesta:2015iho,Giusarma:2016phn,Vagnozzi:2017ovm,Doux:2017tsv,Upadhye:2017hdl,Gualdi:2017iey,Giusarma:2018jei,Gualdi:2018pyw,Loureiro:2018qva,Loureiro:2018pdz,Gualdi:2019ybt}, with some of these works analyzing FS measurements in compressed form), especially in the past months within the context of analyses modelling FS measurements with the effective field theory of LSS (see e.g.~\cite{DAmico:2019fhj,Ivanov:2019pdj,Colas:2019ret,Ivanov:2019hqk,Philcox:2020vvt,DAmico:2020kxu,Nishimichi:2020tvu,Chudaykin:2020aoj,Ivanov:2020ril,DAmico:2020ods,Philcox:2020xbv,Niedermann:2020qbw,Philcox:2020zyp,Chudaykin:2020ghx,Smith:2020rxx}). However, even within the linear or at most weakly non-linear regime, FS measurements contain valuable information and can have their say in the spatial curvature debate. This had already been strongly appreciated back in 2011 in the context of Data Release 7 of the Sloan Digital Sky Survey (SDSS)~\cite{Montesano:2011bp}.

In this work we shall therefore explore what role full-shape galaxy power spectrum measurements take  in the debate around the geometry of the Universe. We shall consider the FS galaxy power spectrum as measured from the CMASS sample of the SDSS-III Baryon Oscillation Spectroscopic Survey (BOSS) Data Release 12 (DR12)~\cite{Alam:2016hwk}. This sample, measuring the clustering of massive galaxies at an effective redshift of $z=0.57$, is the largest high-redshift spectroscopic galaxy sample to date. We will find that combining this FS measurement with P18 data also appears to indicate a spatially flat Universe, in the same way that BAO measurements do. However, much as with BAO data, we will find FS data to be in tension with P18 data within the assumption of a curved Universe. Therefore, another goal of our paper will be that of assessing the level of discordance between P18 and FS data when moving beyond a spatially flat Universe.

The rest of this paper is then organized as follows. In Section~\ref{sec:curvature}, we begin by introducing our notation and providing background information on the role of spatial curvature in cosmological observations, focusing on CMB and LSS measurements. In Section~\ref{sec:methods} we  discuss the methods and datasets we adopt in our study. Our results are presented in Section~\ref{sec:results}, and in particular in Section~\ref{subsec:consistency} we discuss the consistency between the \textit{Planck} and FS galaxy power spectrum datasets when assuming a curved Universe, while in Section~\ref{subsec:other} we further investigate the implications of our previous findings, making use of additional probes which we combine with \textit{Planck}. Finally, in Section~\ref{sec:conclusions} we provide concluding remarks. Technical details concerning the FS galaxy power spectrum modelling and likelihood are given in~\ref{sec:appendix}.

\section{The role of spatial curvature in cosmological observations}
\label{sec:curvature}

We work under the assumption of the cosmological principle, according to which the Universe is homogeneous and isotropic on large scales, and of General Relativity. Under these assumptions, and working in the usual reduced spherical polar coordinates $(r,\theta,\phi)$, the Universe is described by the Friedmann-Lema\^{i}tre-Robertson-Walker metric, with line element:
\begin{eqnarray}
ds^2 = -dt^2+a^2(t) \left [ \frac{dr^2}{1-Kr^2} + r^2 \left ( d\theta^2+\sin^2\theta d\phi^2 \right ) \right ]\,,
\label{eq:flrw}
\end{eqnarray}
where $t$ denotes cosmic time and the scale factor $a$ characterizes the expansion or contraction across cosmic time of homogeneous and isotropic spatial slices. The parameter $K$ characterizes the constant spatial curvature of the spatial slices, with $K=0$ corresponding to flat Euclidean space, positive spatial curvature $K>0$ to closed hyperspherical space, and negative spatial curvature $K<0$ to open hyperbolic space. At the level of the Friedmann equations, curvature gives an effective fractional contribution to the energy budget quantified through the so-called curvature parameter $\Omega_K\equiv-K/(H a)^2$, where $H$ is the Hubble factor. Note that $\Omega_K$ comes with opposite sign with respect to $K$.

When considering only measurements of the CMB temperature anisotropy power spectrum (excluding ultra-large scales which are dominated by cosmic variance), and considering only primary anisotropies, it is not possible to place strong constraints on the curvature parameter $\Omega_K$. The reason is the well-known geometrical degeneracy~\cite{Bond:1997wr,Zaldarriaga:1997ch,Efstathiou:1998xx}, \textit{i.e.}\ the fact that various combinations of cosmological parameters can lead to the same value of the angular diameter distance to last-scattering, and hence to the same angular scale for the first peak of the CMB power spectrum $\theta_s$, assuming no changes to early Universe physics which would affect the sound horizon at last-scattering $r_s$. The result is that all these combinations of cosmological parameters lead to approximately the same CMB power spectrum. In other words, there are various combinations of the matter density parameter $\Omega_m$, curvature density parameter $\Omega_K$, and Hubble constant $H_0$, which have identical CMB spectra as that of a spatially flat model with $\Omega_K=0$. It is important to note that the CMB can constrain quite well the physical matter density $\omega_m \equiv \Omega_mh^2$, both through the so-called ``potential envelope'' or early integrated Sachs-Wolfe (EISW) effect, and the lensing-induced smoothing of the higher acoustic peaks.

Because of the direction of the mutual $\Omega_m$-$\Omega_K$-$H_0$ degeneracies, once all parameters are marginalized over, the $\Omega_K$ posterior might be skewed towards negative values, a result known since the time of BOOMERanG~\cite{Melchiorri:2000px}. Nonetheless, using simulated data, it was shown in dV19 that a \textit{Planck}-like experiment should be able to constrain $\Omega_K$ to $2\%$ without substantial bias towards closed models. This is mainly due to the effect of gravitational lensing at small angular scales (now strongly constrained by \textit{Planck}) that helps break the geometrical degeneracy. This has been beautifully demonstrated by the ACT collaboration through the ACT-DR4+WMAP dataset combination, from which one gets $\Omega_K=-0.001^{+0.014}_{-0.010}$ at 68\%~confidence level (C.L.)~\cite{Aiola:2020azj}. A crucial point here is therefore that measurements of the CMB angular power spectra are currently the \textit{only} observable that can in principle significantly constrain the curvature of the universe independently from any external datasets.

In any case, it is desirable to combine CMB measurements with additional late-time measurements which can further help in breaking the geometrical degeneracy. One example consists of late-time BAO distance and expansion rate measurements, which help to nail down $\Omega_m$ and $H_0$ (especially ruling out low values of $H_0$ around $\sim 50$ km/s/Mpc, in strong tension with local measurements) and hence considerably improve the determination of $\Omega_K$. This has already been noticed in many works, especially in the recent EG20~\cite{Efstathiou:2020wem}. In this work, we shall explore the power of full-shape (FS) galaxy power spectrum measurements. Despite FS and BAO measurements come from the same galaxy survey, they are based on quite different types of analyses and suffer from completely different types of systematics. Therefore, it is worthwhile, timely, and interesting to consider what FS measurements have got to add in the recent debate concerning the spatial curvature of the Universe and more generally cosmic concordance. Related analyses have been performed by the BOSS collaboration making use of configuration space and Fourier space clustering wedges in~\cite{Sanchez:2016sas,Grieb:2016uuo}, by the DES collaboration through a joint $3\times2$pt analysis of galaxy clustering and weak lensing data in~\cite{Abbott:2018xao}, by the eBOSS collaboration in~\cite{Alam:2020sor}, and from an independent re-analysis in~\cite{Chudaykin:2020ghx}.

It is worth briefly discussing what one gains by adding FS data to CMB measurements (for two recent very complete discussions see for instance~\cite{Ivanov:2019pdj,Chudaykin:2020ghx}). The relative height of odd and even peaks in the CMB depends on the physical baryon density parameter $\omega_b \equiv \Omega_bh^2$, with $h$ the dimensionless Hubble parameter. On the other hand, with $\omega_c \equiv \Omega_ch^2$ the physical dark matter (DM) density parameter, the relative amplitude of the BAO wiggles in the FS measurements, as well as the small-scale baryon-induced suppression therein, both help improving the determination of $\omega_b/\omega_c$.

Another important source of information in FS measurements is the turnaround scale: $P(k)$ exhibits a break at $k_{\rm eq}$, the horizon wavenumber at matter-radiation equality: modes entering the horizon before matter-radiation equality are suppressed by radiation pressure, unlike modes entering afterwards. The horizon wavenumber at matter-radiation equality scales as $k_{\rm eq} \propto \Omega_mh^2$. However, since one usually works in redshift space (assuming a fiducial cosmology), observable wavenumbers are typically quoted in units of $h\,{\rm Mpc}^{-1}$, which implies that the quantity governing FS measurements is not $\Omega_mh^2$, but actually $\Omega_mh$ (see~\cite{Efstathiou:1992sy,Ma:1996za,Efstathiou:2001cw,Tegmark:2006az,Reid:2009xm} for early discussions). This quantity is typically referred to as the ``shape parameter'' and denoted by $\Gamma \equiv \Omega_mh$.

As we already discussed, CMB measurements can constrain $\Omega_mh^2$ through the EISW and lensing effects. A simultaneous measurement of $\Omega_mh$ from FS measurements therefore can enormously help in breaking the degeneracy between $\Omega_m$ and $H_0$ and better determining each of these two quantities. The result is that the geometrical degeneracy present with CMB data alone can be alleviated by the inclusion of FS measurements: this important point had already been appreciated in early work~\cite{Ferramacho:2008ap,Montesano:2011bp}.

So far, this discussion only exploited shape information. However, FS measurements also contain geometric information, as well explained in~\cite{Ivanov:2019pdj}. In fact, the position of the BAO wiggles in momentum space depends on the ratio $r_d/D_V$, with $r_d$ the sound horizon at baryon drag and $D_V$ the volume-averaged distance to the effective redshift of the galaxy sample. Once $\omega_b$ and $\omega_c$ are known as discussed above, $r_d$ is also known, and hence $D_V$ can be inferred. Within the minimal $\Lambda$CDM model, $H_0$ can then be tuned to match the inferred value of $D_V$ and by extension the location of baryonic features in FS measurements (see e.g.~\cite{Ivanov:2019pdj}). In addition, the overall amplitude of the power spectrum depends on $\Omega_m$ (see e.g. Chapter~6.1 of~\cite{Lesgourgues:2018ncw}), and thus improves the determination of $H_0$, since $\Omega_m=(\omega_b+\omega_c)/h^2$, with $\omega_b$ and $\omega_c$ known.

The considerations made so far on the geometrical degeneracy being broken by the combination of CMB and FS measurements were made for the minimal spatially flat $\Lambda$CDM model. However, our considerations  extend to models with non-zero $\Omega_K$ as well. Indeed, FS measurements help, once more, in resolving the mutual $\Omega_m$-$\Omega_K$-$H_0$ degeneracies present with CMB data alone. In particular, FS information allows us to exclude low values of $H_0$ (in strong tension with local measurements), which CMB data alone would otherwise tolerate (see e.g.\ Fig.~1a in EG20), the same way BAO do.

\section{Methodology}
\label{sec:methods}

The cosmological observations we shall consider in the following are:
\begin{itemize}
\item Measurements of CMB temperature and polarization anisotropies, as well as their cross-correlation, from the \textit{Planck} 2018 legacy data release~\cite{Akrami:2018vks,Aghanim:2018eyx,Aghanim:2019ame}. This combination is referred to as \textit{Planck TTTEEE+lowE} in the \textit{Planck} papers, and combines the \texttt{plik\_rd12\_HM\_v22b\_TTTEEE}, \texttt{simall\_100$\times$143\_offlike5\_EE\_Aplanck\_B}, and \texttt{commander\_dx12\_v3\_2\_29} likelihoods. We refer to this dataset as \textbf{\textit{Planck}} (and occasionally we shall refer to it as P18).
\item Measurement of the angle-averaged (monopole moment) full-shape power spectrum of the BOSS DR12 CMASS sample at an effective redshift $z_{\rm eff}=0.57$ as measured in~\cite{Gil-Marin:2015sqa}. The theoretical modelling and likelihood of this measurement (including the way we model survey geometry effects) is described in more detail in Appendix~\ref{sec:appendix}. We analyze the measurements within the wavenumber range $0.03\,h\,{\rm Mpc}^{-1}<k<0.135\,h\,{\rm Mpc}^{-1}$, to minimize the impact of observational systematics on large scales, and theoretical systematics (non-linearities) on small scales. We refer to this dataset as \textbf{\textit{FS}}.~\footnote{We remark that this is not the consensus full-shape power spectrum measured at the three optimally binned effective redshifts $z_{\rm eff}=0.38,0.51,0.61$ of the combined sample by the BOSS collaboration in~\cite{Alam:2016hwk}, but the full-shape power spectrum measured in the earlier Gil-Mar\'{i}n \textit{et al.} 2016~\cite{Gil-Marin:2015sqa} BOSS paper, which still adopted the ``traditional'' LOWZ/CMASS splitting of the galaxy sample. In~\cite{Alam:2016hwk}, it was shown that cosmological constraints obtained from~\cite{Gil-Marin:2015sqa} and~\cite{Alam:2016hwk} are in good agreement.} For simplicity, we have not included measurements of the quadrupole moment, although we note that adding the latter would lead to tighter constraints than those obtained from the monopole alone.
\item BAO distance measurements from the 6dFGS~\cite{Beutler:2011hx}, SDSS-MGS~\cite{Ross:2014qpa}, and BOSS DR12~\cite{Alam:2016hwk} surveys. We refer to this dataset as \textbf{\textit{BAO}}. Note that the former two measurements constrain the ratio of the volume distance at the effective redshift of the galaxy sample $z_{\rm eff}$ to the sound horizon $D_V(z_{\rm eff})/r_s$, whereas the latter constrain separately the ratio of the angular diameter distance to the sound horizon $D_A(z_{\rm eff})/r_s$, and the product of the Hubble rate and the sound horizon $H(z_{\rm eff})r_s$.
\end{itemize}
The main dataset combination we consider is \textit{Planck}+\textit{FS}. To compare our results with those of H19, dV19, and EG20, we also consider the \textit{Planck}+\textit{BAO} dataset combination. On the other hand, we do not combine the \textit{FS} and \textit{BAO} datasets, given the strong correlation between the \textit{FS} dataset and the two high-redshift bins of the BOSS DR12 BAO measurements. Note that we do not consider measurements of the CMB lensing power spectrum from \textit{Planck}, as reconstructed from the temperature 4-point function. The reason is that we only are interested in seeing whether the preference for a closed Universe from \textit{Planck} temperature and polarization anisotropies alone survives once LSS data, in the form of either BAO or FS measurements, are included.

At a later stage, to further investigate the significance of the results we obtain within the \textit{Planck}+\textit{FS} and \textit{Planck}+\textit{BAO} dataset combinations, we shall consider two further datasets:
\begin{itemize}
\item Uncalibrated magnitude-redshift relation of Hubble flow SNeIa from the \textit{Pantheon} sample, consisting of distance moduli measurements for 1048 SNeIa in the redshift range $0.01<z<2.3$~\cite{Scolnic:2017caz}. We refer to this dataset as \textbf{\textit{Pantheon}}.
\item 31 cosmic chronometer measurements of $H(z)$, from the differential age evolution of massive, early-time, passively evolving galaxies~\cite{Jimenez:2001gg}, in the range $0.07<z<1.965$, compiled in~\cite{Jimenez:2003iv,Simon:2004tf,Stern:2009ep,Moresco:2012by,Moresco:2015cya,Moresco:2016mzx,Ratsimbazafy:2017vga} (see e.g. Tab.~1 in~\cite{Vagnozzi:2020dfn} for a summary of these measurements). We refer to this dataset as \textbf{\textit{CC}}.
\end{itemize}

Model-wise, we consider a one-parameter extension of the concordance $\Lambda$CDM model, for a total of 7 cosmological parameters: the baryon and cold DM physical densities $\Omega_bh^2$ and $\Omega_ch^2$, the angular size of the sound horizon at last-scattering $\theta_s$, the optical depth to reionization $\tau$, the amplitude and tilt of the primordial scalar power spectrum $A_s$ and $n_s$, and the curvature parameter $\Omega_K$. We refer to this seven parameter model as $K\Lambda$CDM, and adopt uniform priors on all seven parameters unless otherwise specified. In particular, we vary $\Omega_K$ within the range $\Omega_K \in [-0.3,0.3]$, as done by the \textit{Planck} collaboration~\cite{Aghanim:2018eyx} and in the dV19 and H19 papers. Moreover, despite an uniform prior on $\Omega_K$ not necessarily being highly motivated from e.g.\ inflation (see the discussion in EG20), we note that here we are providing observational constraints on the value of $\Omega_K$, detached from any underlying theoretical model, and that an inflationary prior that strongly prefers a flat Universe could introduce some amount of bias in our results. We note that \texttt{CosmoMC} imposes an implicit prior on $H_0$ within the range $[20;100]\,{\rm km}\,{\rm s}^{-1}\,{\rm Mpc}^{-1}$.

Theoretical predictions for the CMB and galaxy power spectra are obtained using the Boltzmann solver \texttt{CAMB}~\cite{Lewis:1999bs}. We sample the 7-dimensional parameter space by using Monte Carlo Markov Chain (MCMC) methods. Our MCMC chains are generated through a suitably modified version of the cosmological sampler \texttt{CosmoMC}~\cite{Lewis:2002ah}, to which we have included the FS likelihood. We monitor the convergence of the generated chains by using the Gelman-Rubin parameter $R-1$~\cite{Gelman:1992zz}.

One further comment regarding the primordial power spectrum in the presence of spatial curvature is in order before moving forward. It is worth noting that in the presence of spatial curvature characterized by the curvature parameter $K$, \texttt{CAMB} parametrizes the (dimensionless) primordial power spectrum of scalar fluctuations as:
\begin{eqnarray}
\Delta (k) = \frac{(q^2-4K)^2}{q(q^2-K)}k^{n_s-1}\,,
\label{eq:power}
\end{eqnarray}
where $q=\sqrt{k^2+K}$. In this form, Eq.~(\ref{eq:power}) makes a rather specific assumption about how primordial fluctuations extend to scales larger than the curvature scale. The particular functional form chosen ensures that potential fluctuations are constant per logarithmic interval in wavenumber $k$~\cite{White:1995ic}. In the absence of a well agreed upon model for the origin of fluctuations in a curved Universe, how the concept of scale-invariant fluctuations should be generalized to scales close to the curvature scale is not obvious (see e.g.~\cite{Efstathiou:2003hk} for further discussions). While this could have an important impact on the use of the \texttt{commander\_dx12\_v3\_2\_29} likelihood, and in particular its $\chi^2$, we do not expect it to affect our \textit{FS} results significantly, given the much smaller scales we are considering.

For recent works that carefully compute the primordial power spectrum expected in curved inflating Universes by means of the Mukhanov-Sasaki equation, we invite the reader to consult~\cite{Handley:2019anl,Thavanesan:2020lov}. In particular, in~\cite{Handley:2019anl} it was shown that the largest deviations from the ``traditional'' power spectrum are only expected at very large scales, where the primordial power spectrum is truncated below the curvature scale. While in the following we will assume a dimensionless primordial power spectrum given by Eq.~(\ref{eq:power}), it is worth keeping the caveats concerning this choice in mind, which EG20 takes as a further indication for the fact that the posteriors on $\Omega_K$ should not be over-interpreted. It is important to stress that using more accurate predictions for the primordial power spectrum predicted by inflation in a curved Universe as in~\cite{Handley:2019anl} might increase the evidence for a closed universe from P18 data alone.

Besides providing constraints on cosmological parameters (most importantly on $\Omega_K$) from the \textit{Planck}+\textit{FS} dataset combination, another important goal of ours is to assess the consistency of this dataset combination, within the assumption of a $K\Lambda$CDM Universe. Should these two datasets be found to be in tension, their combination should be viewed with caution, regardless of any ability to break the geometrical degeneracy, as pointed out in H19 and dV19. A first rough but informative step towards assessing the concordance between \textit{Planck} and \textit{FS} assuming a curved Universe is to estimate by how much the best-fit $\chi^2$ \textit{increases} when adding \textit{FS} to \textit{Planck} within a $K\Lambda$CDM model, and compare this $\Delta \chi^2$ to the same quantity obtained assuming $\Lambda$CDM. Or, at a fixed dataset combination, to estimate by how much the best-fit $\chi^2$ \textit{decreases} for $K\Lambda$CDM relative to the baseline $\Lambda$CDM model (the $\chi^2$ cannot increase as $\Lambda$CDM is nested within $K\Lambda$CDM).

Besides looking at the $\Delta \chi^2$, more robust concordance/discordance diagnostics exist in the literature (see e.g.~\cite{Karpenka:2014moa,MacCrann:2014wfa,Lin:2017ikq,Lin:2017bhs,Adhikari:2018wnk,Raveri:2018wln,Nicola:2018rcd,Handley:2019wlz,Handley:2019pqx,Garcia-Quintero:2019cgt,Lemos:2019txn,Raveri:2019gdp}). To assess the consistency between \textit{Planck} and \textit{FS}, we follow the method outlined in~\cite{Joudaki:2016mvz,Hildebrandt:2016iqg,Joudaki:2016kym} and utilized in dV19. This method makes use of the so-called deviance information criterion (DIC). The DIC is a model comparison tool, whose definition is grounded in information theory, and is given by~\cite{Spiegelhalter:2002ghw}:
\begin{eqnarray}
{\rm DIC} = \chi^2(\hat{\theta})+2p_D\,.
\label{eq:dic}
\end{eqnarray}
In Eq.~(\ref{eq:dic}), $\chi^2(\hat{\theta})=-2\ln {\cal L}_{\rm max}$ is the best-fit effective $\chi^2$, which is thus evaluated at the value of the parameter vector $\hat{\theta}$ yielding the maximum likelihood ${\cal L}_{\rm max}$. Still in Eq.~(\ref{eq:dic}), $p_D$ is the Bayesian complexity factor, which acts to penalize more complex models which do not yield a sufficient improvement in fit, and is given by:
\begin{eqnarray}
p_D = \overline{\chi^2(\theta)}-\chi^2(\hat{\theta})\,,
\label{eq:pd}
\end{eqnarray}
with $\overline{\chi^2(\theta)}$ denoting an average of the effective $\chi^2$ over the posterior distribution. In comparing an extended model to a reference model (e.g.\ $\Lambda$CDM), negative values of $\Delta$DIC indicate that the extended model is favored (in a model comparison sense).

In this work, rather than using the DIC as a model comparison tool, we will use a DIC-grounded statistic to estimate the concordance between two datasets $D_1$ and $D_2$, with the underlying cosmological model being fixed. The DIC-grounded statistic ${\cal I}$ we make use of is given by the following expression~\cite{Joudaki:2016mvz}:
\begin{eqnarray}
{\cal I}(D_1\,,D_2) \equiv \exp \left [ -\frac{{\cal G}(D_1\,,D_2)}{2} \right ] \,,
\label{eq:id1d2}
\end{eqnarray}
where the quantity ${\cal G}(D_1\,,D_2)$ is given by:
\begin{eqnarray}
{\cal G}(D_1\,,D_2) \equiv {\rm DIC}(D_1 \cup D_2)-{\rm DIC}(D_1)-{\rm DIC}(D_2)\,,
\label{eq:gd1d2}
\end{eqnarray}
with ${\rm DIC}(D_1 \cup D_2)$ indicating the deviance information criterion evaluated from the joint dataset consisting of the combination of $D_1$ and $D_2$.

We use ${\cal I}$ to estimate the level of concordance or discordance between the \textit{Planck} and \textit{FS} datasets. In particular, a positive value of $\log_{10}{\cal I}$ indicates agreement between the two datasets, and conversely for a negative value of $\log_{10}{\cal I}$. We qualify the level of concordance or discordance between \textit{Planck} and \textit{FS} using the Jeffreys-like scale used in~\cite{Joudaki:2016mvz}. If $\log_{10}{\cal I}<0$, the level of discordance between $D_1$ and $D_2$ is considered ``substantial'' if $\vert \log_{10}{\cal I} \vert >0.5$, ``strong'' if $\vert \log_{10}{\cal I} \vert >1.0$, and ``decisive'' if $\vert \log_{10}{\cal I} \vert >2.0$, whereas a value of $\vert \log_{10}{\cal I} \vert <0.5$ indicates no significant level of discordance.

\section{Results}
\label{sec:results}

\begin{table*}
\begin{center}                              
\scalebox{1.0}{
\begin{tabular}{|c|ccc|}       
\hline\hline
\backslashbox{Parameters}{Dataset} & \textit{Planck} & \textit{Planck}+\textit{BAO} & \textit{Planck}+\textit{FS} \\ \hline
$\Omega_K$ & $-0.044^{+0.018}_{-0.015}$ & $0.0008 \pm 0.0019$ & $0.0023 \pm 0.0028$ \\
$H_0\,[{\rm km/s/Mpc}]$ & $54.36^{+3.25}_{-3.96}$ & $67.88 \pm 0.66$ & $68.59^{+1.08}_{-1.20}$ \\
$\Omega_m$ & $0.485^{+0.058}_{-0.068}$ & $0.310 \pm 0.007$ & $0.304 \pm 0.010$ \\ \hline
$\Delta \chi^2$ & $-10.9$ & $-0.6$ &  $-1.0$ \\
\hline\hline                                                  
\end{tabular}}
\end{center}
\caption{68\%~C.L. constraints on selected cosmological parameters ($\Omega_K$, $H_0$, and $\Omega_m$) within the seven-parameter $K\Lambda$CDM model. The final row reports the $\Delta \chi^2$ with respect to the 6-parameter $\Lambda$CDM model for the same dataset combination.
}
\label{tab:parameters}                                              
\end{table*}       

Cosmological constraints on the curvature density parameter $\Omega_K$, the Hubble constant $H_0$, and the matter density parameter $\Omega_m$, obtained within the seven-parameter $K\Lambda$CDM model, are reported in Table~\ref{tab:parameters}. We first consider the \textit{Planck} dataset alone, and then in combination with the \textit{BAO} and \textit{FS} datasets, one at a time. For each of these three dataset combinations (\textit{Planck}, \textit{Planck}+\textit{BAO}, and \textit{Planck}+\textit{FS}), in Table~\ref{tab:parameters} we also report the $\Delta \chi^2$, the difference in the best-fit $\chi^2$ for the $K\Lambda$CDM model with respect to the six-parameter $\Lambda$CDM model for the same dataset combination.

Notice from the first column of Table~\ref{tab:parameters}, that for the \textit{Planck}-only case we recover the well-known $\Omega_K=-0.044^{+0.018}_{-0.015}$ at 68\%~C.L., with a substantial $\vert \Delta \chi^2 \vert>10$ improvement in the fit with respect to the $\Lambda$CDM case. Within a non-flat Universe, \textit{Planck} data alone also prefers a substantially lower value of $H_0$ (in strong tension with local measurements) and a significantly higher value of $\Omega_m$ (in strong tension with independent LSS measurements), reflecting the direction of the aforementioned $\Omega_m$-$\Omega_K$-$H_0$ geometrical degeneracy~\cite{Bond:1997wr,Zaldarriaga:1997ch,Efstathiou:1998xx}. The second column of Table~\ref{tab:parameters} reports the other well-known result that combining \textit{Planck} and \textit{BAO} suggests once more a spatially flat Universe, with $\Omega_K=0.0008 \pm 0.0019$, whereas $H_0$ and $\Omega_m$ move towards values which are in agreement with independent late-time probes. For the \textit{Planck}+\textit{BAO} dataset combination, the improvement in fit with respect to $\Lambda$CDM is extremely mild, with a $\Delta \chi^2=-0.6$.

Before moving forward, it is worth recalling where part of the P18 preference for a closed Universe is coming from. As explained by the Planck collaboration~\cite{Aghanim:2018eyx}, by dV19, and by EG20, this preference is partially driven by the anomalous preference of the \textit{Planck} temperature anisotropy power spectrum for a higher amount of lensing. In other words, the acoustic peaks in temperature are slightly more smoothed than one would expect within the baseline $\Lambda$CDM model given the other cosmological parameters, an effect which one could easily be tempted to interpret as a lensing excess. This anomaly is quantified by the phenomenological parameter $A_L$~\cite{Calabrese:2008rt}, which rescales the lensing amplitude in the CMB power spectra: in particular, \textit{Planck} data appears to prefer $A_L>1$, with a preference of about $2.8\sigma$. It is unclear whether the lensing anomaly is a true anomaly or a statistical fluctuation, although a re-analysis of \textit{Planck} High Frequency maps with access to a larger sky fraction, but also with the removal of the $100 \times 100$ GHz spectrum, appears to support the latter interpretation~\cite{Efstathiou:2019mdh}.~\footnote{Proposed explanations for the lensing anomaly include modified gravity~\cite{DiValentino:2015bja,DiValentino:2020evt}, compensated isocurvature perturbations~\cite{Munoz:2015fdv,Valiviita:2017fbx,Smith:2017ndr}, and oscillations in the primordial power spectrum~\cite{Domenech:2019cyh}, possibly produced during an early period alternative to inflation~\cite{Domenech:2020qay}.} The same interpretation is also supported by the latest ACT results, which are consistent with $A_L=1$~\cite{Aiola:2020azj}. Tt is worth keeping in mind that the \textit{Planck} preference for $\Omega_K<0$ is partially driven by this anomaly, as a closed Universe can naturally accommodate a higher $\Omega_m$ and hence a higher $A_L$, besides providing a slightly better fit to a number of anomalously low features in the low-$\ell$ multipoles of the CMB temperature power spectrum (see e.g.~\cite{Efstathiou:2003hk}).

Returning to the main topic of this work, namely constraints on $\Omega_K$ from FS measurements, the new results of this paper are those shown in the third column of Table~\ref{tab:parameters}. Note that replacing the \textit{BAO} dataset with the \textit{FS} one, which also allows to break the geometrical degeneracy, leads to qualitatively similar results: the \textit{Planck}+\textit{FS} dataset combination also indicates a spatially flat Universe to sub-percent precision, with $\Omega_K=0.0023 \pm 0.0028$. The values of $H_0$ and $\Omega_m$ inferred are also in much better agreement with independent late-time probes, with $H_0=68.6 \pm 1.2\,{\rm km}/{\rm s}/{\rm Mpc}$ and $\Omega_m=0.304 \pm 0.010$. The improvement in the $\chi^2$ with respect to $\Lambda$CDM is still very mild, with $\Delta \chi^2=-1.0$. These results already allow us to provide an answer to the question posed in the introductory part of this paper: does the FS galaxy power spectrum also indicate a spatially flat Universe once combined with P18? The answer, as we see from the third column of Table~\ref{tab:parameters}, is \textit{yes}. While one could perhaps have expected this to be the case a priori, we believe this answer is actually rather non-trivial, and serves as a strong and robust consistency analysis between the \textit{BAO} and the \textit{FS} techniques when dealing with LSS observations. Note indeed that the \textit{Planck}+\textit{BAO} and \textit{Planck}+\textit{FS} dataset combinations are in good agreement (within better than $1\sigma$) as far as the central values of the cosmological parameters are concerned, with the latter preferring slightly higher values of $H_0$ and slightly lower values of $\Omega_m$.  Particularly worth noting is that the \textit{Planck}+\textit{FS} dataset combination is slightly less constraining than the \textit{Planck}+\textit{BAO} one. The uncertainties on $\Omega_K$, $H_0$, and $\Omega_m$ are respectively $\approx 50\%$, $\approx 70\%$, and $\approx 40\%$ larger for \textit{Planck}+\textit{FS} compared to \textit{Planck}+\textit{BAO}.

While the previous result might at first glance seem unexpected, it is actually in agreement with earlier studies on the subject, such as~\cite{Hamann:2010pw,Vagnozzi:2017ovm,Ivanov:2019hqk}. First of all, it is worth reminding ourselves the obvious point that our \textit{BAO} dataset includes also measurements from the 6dFGS and SDSS-MGS surveys, alongside the lowest redshift bin from BOSS DR12, whereas our \textit{FS} measurements only include the power spectrum from the BOSS DR12 CMASS sample. Secondly, we have made us of pre-reconstruction FS measurements, whereas BAO datasets are typically based on post-reconstruction measurements, where the reconstruction procedure is performed to enhance the BAO signal-to-noise~\cite{Eisenstein:2006nk}, and includes additional non-linear information which are are instead not accessing within the FS measurements. Thirdly, the BOSS collaboration has measured the BAO feature in both the transverse and parallel directions, obtaining separate constraints on the angular diameter distance $D_A(z_{\rm eff})$ and Hubble rate $H(z_{\rm eff})$ at the effective redshift of the galaxy sample $z_{\rm eff}$. The measurements of the FS monopole we have used are instead only sensitive to the volume distance $D_V(z_{\rm eff})$, much like the earlier spherically averaged BAO measurements. Furthermore, the modelling of the FS measurements requires several additional nuisance parameters, which further degrade the obtained constraints once they are marginalized over. Additionally, it has been recently argued in~\cite{Ivanov:2019hqk} that the fact that \textit{FS} and \textit{BAO} measurements lead to similar error bars, with the latter being slightly more constraining, is simply a coincidence given the current BOSS volume and BAO reconstruction efficiency. In particular, even for ideal BAO reconstruction, within future galaxy surveys covering larger volumes, the \textit{FS} information would eventually be expected to supersede the \textit{BAO} one.

\begin{figure}[!t]
\centering
\includegraphics[width=0.8\linewidth]{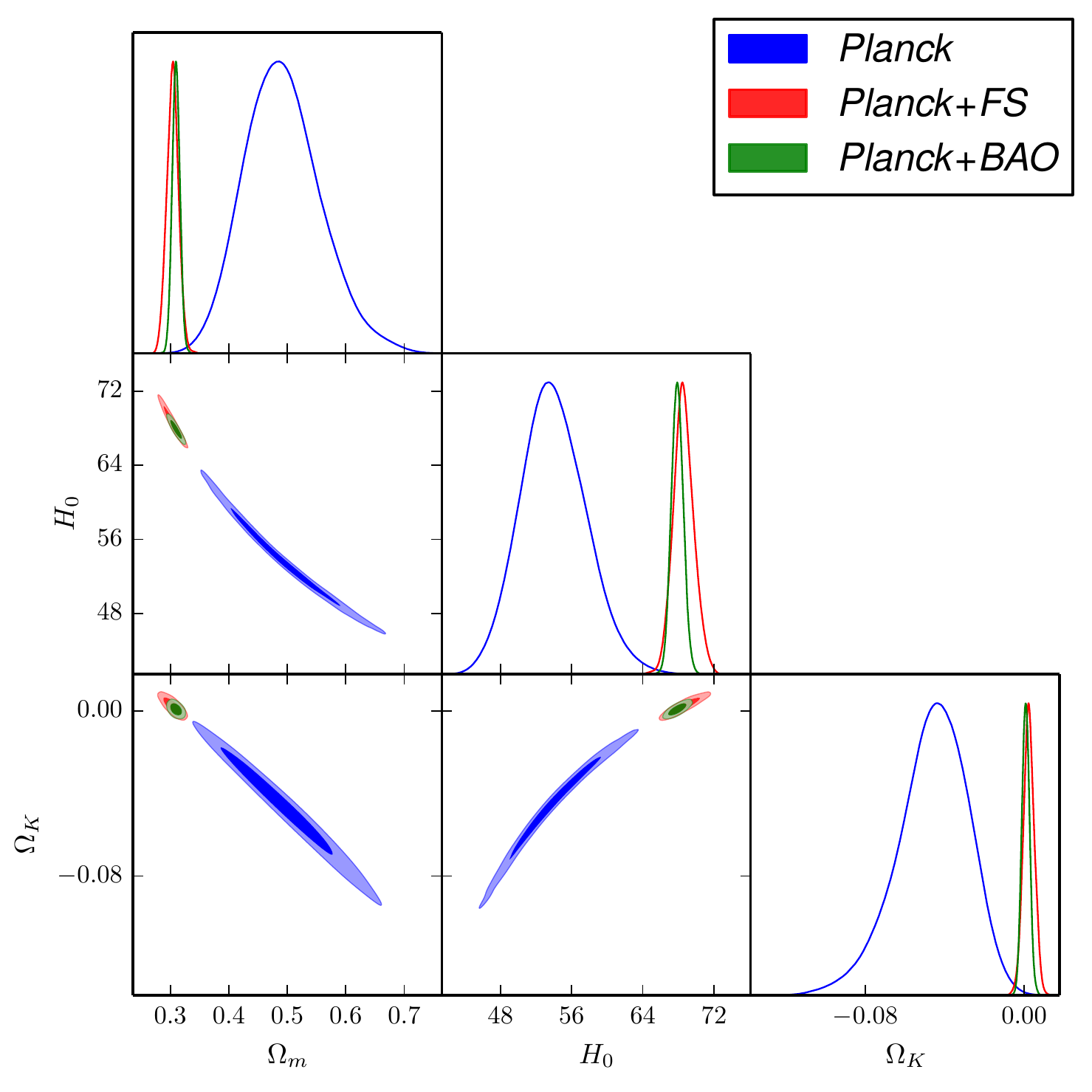}
\caption{Triangular plot showing 2D joint and 1D marginalized posterior probability distributions for $\Omega_m$, $H_0$, and $\Omega_K$ from the \textit{Planck} (blue contours), \textit{Planck}+\textit{BAO} (green contours), and \textit{Planck}+\textit{FS} (red contours) dataset combinations. This figure provides a visual representation of the results reported in Table~\ref{tab:parameters}, and constitutes the most important result of this work.}
\label{fig:base_omegak_tri}
\end{figure}

A visual representation of our results is given in the triangular plot in Fig.~\ref{fig:base_omegak_tri}, where we show the constraints on $\Omega_m$, $H_0$, and $\Omega_K$, for \textit{Planck} (blue contours), \textit{Planck}+\textit{BAO} (green contours), and \textit{Planck}+\textit{FS} (red contours). From Fig.~\ref{fig:base_omegak_tri} we visually see: 1) the fact that both the \textit{BAO} and \textit{FS} datasets pull the \textit{Planck}-only results back towards a spatially flat Universe; 2) the good overall agreement between \textit{Planck}+\textit{BAO} and \textit{Planck}+\textit{FS}; and 3) the slightly weaker constraining power of \textit{FS} as opposed to \textit{BAO}.

\begin{table*}[!b]
\begin{center} 
\scalebox{1.0}{
\begin{tabular}{|c|cc|}       
\hline\hline                                                                  
\backslashbox{Model}{Dataset} & \textit{Planck}+\textit{BAO} & \textit{Planck}+\textit{FS} \\ \hline
$\Lambda$CDM & $+6.1$ & $+22.0$ \\
$K\Lambda$CDM & $+16.8$ & $+31.9$ \\
\hline\hline                                                  
\end{tabular}}
\end{center}
\caption{$\Delta \chi^2$ with respect to the \textit{Planck}-only dataset combination, for both the \textit{Planck}+\textit{BAO} and \textit{Planck}+\textit{FS} dataset combinations, within the $\Lambda$CDM and $K\Lambda$CDM models.}
\label{tab:chi2}                                              
\end{table*}

\subsection{Tension between \textit{Planck} and \textit{FS} within a curved Universe}
\label{subsec:consistency}

So far, we have seen that the \textit{Planck}+\textit{FS} dataset combination appears to indicate a spatially flat Universe, much as the \textit{Planck}+\textit{BAO} dataset combination. In particular, we have been able to appreciate the pivotal role of the \textit{FS} dataset in breaking the geometrical degeneracy. However, it is still important to assess the level of concordance between \textit{Planck} and \textit{FS} within the context of a curved Universe, as done earlier with other datasets in H19 and dV19. We follow the methodology outlined in Section~\ref{sec:methods}.

First of all, we check by how much the best-fit $\chi^2$ increases when adding either the \textit{BAO} or \textit{FS} dataset to \textit{Planck} within either the $\Lambda$CDM or the $K\Lambda$CDM model. Notice from Table~\ref{tab:chi2} that within the $\Lambda$CDM picture, adding \textit{BAO} to \textit{Planck} leads to an increase of $\Delta \chi^2=+6.1$, whereas adding \textit{FS} to \textit{Planck} leads to an increase of $\Delta \chi^2=+22.0$, consistent with the 20 \textit{FS} datapoints we have considered. These figures suggest no significant tension between either \textit{Planck} and \textit{BAO} or \textit{Planck} and \textit{FS} within the $\Lambda$CDM model, with the former result agreeing with earlier findings in H19 and dV19.

If we instead consider the $K\Lambda$CDM model, we see that adding \textit{BAO} to \textit{Planck} leads to an increase of $\Delta \chi^2=+16.8$, consistent with the earlier findings of H19 and dV19 that the two datasets are in tension with each other within a curved Universe. The situation is qualitatively similar when adding \textit{FS} to \textit{Planck}, in which case we find an increase of $\Delta \chi^2=+31.9$. Therefore, the \textit{FS} and \textit{Planck} datasets appear to be in tension when assuming a curved Universe, as one could have guessed. The corroboration we provide here based on real data is, of course, highly reassuring.

The tension between \textit{Planck} and \textit{FS} within a curved Universe is visually apparent in Fig.~\ref{fig:base_omegak_tri}. There, we clearly see that the $95\%$~C.L. regions for both the \textit{Planck}+\textit{BAO} and \textit{Planck}+\textit{FS} dataset combinations in the $\Omega_m$-$H_0$-$\Omega_K$ plane are well separated from the corresponding contours obtained from \textit{Planck} alone. The discordance is also rather clear in the 1D marginalized posteriors for these three parameters, particularly for $\Omega_K$.

We quantify the degree of discordance between the \textit{Planck} and \textit{FS} datasets within the $K\Lambda$CDM model using the DIC-grounded ${\cal I}$ diagnostic defined in Eq.~(\ref{eq:id1d2}). We find $\log_{10}{\cal I}$(\textit{Planck},\textit{FS})$\approx -2.5$, a decisive tension on the Jeffreys-like scale we adopted. This is visually apparent from the wide separation between the blue and red contours in Fig.~\ref{fig:base_omegak_tri}.

\subsection{Other probes to break the geometrical degeneracy}
\label{subsec:other}

Our previous results indicate that: \textit{i)} the \textit{Planck}+\textit{FS} results are overall in good agreement with the \textit{Planck}+\textit{BAO} results, with \textit{ii)} both combinations pointing towards a spatially flat Universe, but \textit{iii)} at the cost of a strong tension between \textit{Planck} and these external probes within the $K\Lambda$CDM model. Given that the \textit{FS} and \textit{BAO} measurements rely on similar datasets, with analysis techniques differing, the overall consistency between the \textit{Planck}+\textit{FS} and \textit{Planck}+\textit{BAO} results is not surprising, but reassuring. We further note that, while inconsistent within the $K\Lambda$CDM model, \textit{Planck} is consistent with both \textit{FS} and \textit{BAO} within the $\Lambda$CDM model.

This last point warrants further investigation. In fact, besides the fact that the resulting combined constraints should be viewed with caution, an inconsistency between two datasets within a given cosmological model can imply a failure either in one of the datasets (e.g. unaccounted for systematics), or in the model itself. It is admittedly hard to envisage a scenario wherein unaccounted for systematics only lead to inconsistency within the $K\Lambda$CDM model, but not within $\Lambda$CDM. However, given the fact that the \textit{FS} and \textit{BAO} measurements rely on similar datasets, our results are inconclusive in the sense that it is difficult to tell whether, in the context of the previous discussion, it is the data or the model that should be blamed.

\begin{figure}[!t]
\centering
\includegraphics[width=0.8\linewidth]{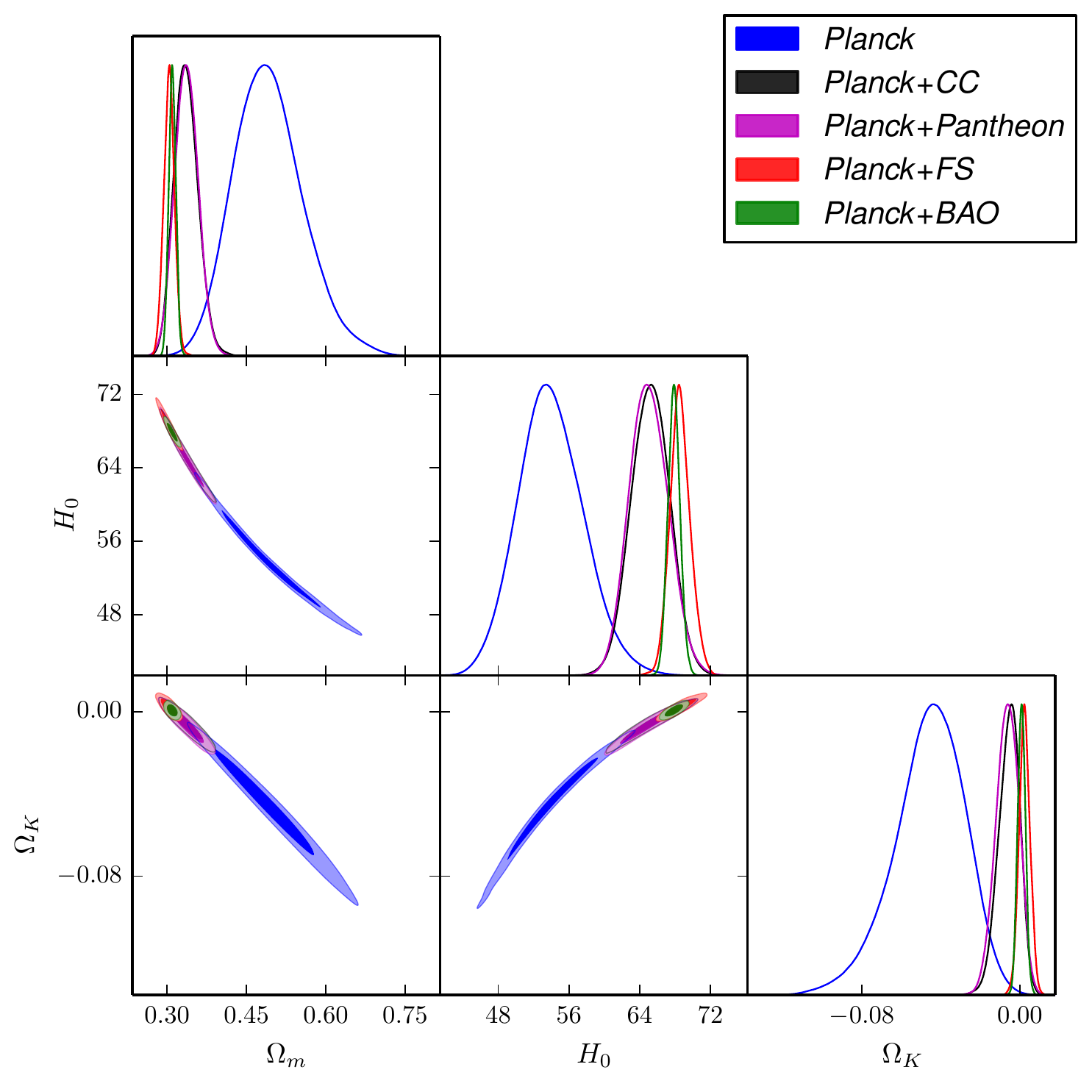}
\caption{As in Fig.~\ref{fig:base_omegak_tri}, but also including the \textit{Planck}+\textit{CC} (black contours) and \textit{Planck}+\textit{Pantheon} (magenta contours) results, which are in better agreement with a spatially flat Universe, while at the same time also exhibiting much milder tensions with \textit{Planck}.}
\label{fig:base_omegak_tri_R1}
\end{figure}

A possible way of addressing this issue could be to use alternative external datasets which help breaking the geometrical degeneracy once combined with \textit{Planck}. The use of the CMB lensing and local $H_0$ measurements has been discussed in~\cite{Handley:2019tkm,DiValentino:2019qzk,DiValentino:2020hov}, and we refer the reader to these works for more information. As discussed in Sec.~\ref{sec:methods}, here we shall consider SNeIa distance moduli from the \textit{Pantheon} dataset, as well as cosmic chronometer measurements of $H(z)$ from the \textit{CC} dataset. The \textit{Pantheon} dataset is sensitive to the (unnormalized) shape of $H(z)$, and hence to its slope, which is directly sensitive to $\Omega_m$: \textit{Pantheon} therefore helps breaking the geometrical degeneracy by virtue of an improved determination of $\Omega_m$. The \textit{CC} dataset, on the other hand, is sensitive to the normalized shape of $H(z)$, and hence helps break the geometrical degeneracy by virtue of improved determinations of both $\Omega_m$ and $H_0$. We note that the role of the \textit{CC} dataset in possibly settling the debate around the shape of the Universe was discussed by one of us in~\cite{Vagnozzi:2020dfn}.

From the \textit{Planck}+\textit{CC} dataset combination, we infer $\Omega_K = -0.0054 \pm 0.0055$, in agreement with the Universe being spatially flat within $\simeq 1\sigma$. We obtain a qualitatively similar results from the \textit{Planck}+\textit{Pantheon} dataset combination, from which we infer $\Omega_K = -0.0064 \pm 0.0058$, a result which is also consistent with the Universe being spatially flat within $\gtrsim 1\sigma$.~\footnote{We note that a method for non-parametrically inferring $\Omega_K$ from the \textit{Pantheon} and \textit{CC} datasets alone was recently proposed by one of us in~\cite{Dhawan:2021mel}, and returns a value of $\Omega_K = -0.03 \pm 0.26$ in excellent agreement with spatial flatness, albeit with substantially larger error bars.} Using the same DIC-grounded statistic discussed earlier, we infer $\log_{10}{\cal I}$(\textit{Planck},\textit{CC})$\approx -0.47$ and $\log_{10}{\cal I}$(\textit{Planck},\textit{Pantheon})$\approx -0.44$. In both cases, the negative value of $\log_{10}{\cal I}$ indicates that \textit{Planck} is in agreement with neither \textit{CC} nor \textit{Pantheon}. The absolute value of $\log_{10}{\cal I}$ is still small enough for the tension to be interpreted as being ``mild'' on the Jeffreys-like scale we adopt, although it is very close to the boundary between ``mild'' and ``definite''. Compared to the \textit{Planck}+\textit{FS} and \textit{Planck}+\textit{BAO} cases, the tension is milder because of both the enlarged error bars (larger by up to a factor of $3$), as well as the central value of the inferred $\Omega_K$ moving towards negative values, more in line with the result from \textit{Planck} alone. A visual representation of these results is given in Fig.~\ref{fig:base_omegak_tri_R1}. From this we clearly see that the \textit{Planck}+\textit{CC}/\textit{Planck}+\textit{Pantheon} constraints are somewhat intermediate between the \textit{Planck} and \textit{Planck}+\textit{FS}/\textit{Planck}+\textit{BAO} ones, with the more negative central values and larger error bars easing the tension with \textit{Planck}, while still being overall consistent with spatial flatness.

These results teach us two important lessons: \textit{i)} it is indeed possible to break the geometrical degeneracy by combining \textit{Planck} with external probes within the $K\Lambda$CDM model and infer a value of $\Omega_K$ consistent with spatial flatness within $\simeq 1\sigma$, but \textit{ii)} this can never be achieved without some amount of tension between \textit{Planck} and these external probes: the best one can hope for is a mild amount of tension, as for the case we have just studied using \textit{CC} and \textit{Pantheon} as external probes. We expect \textit{ii)} to apply more generally than just to the external probes considered here. Consider a generic external dataset \textit{ext} which helps breaking the geometrical degeneracy once combined with \textit{Planck} primary CMB data. Because the latter on their own appear to prefer a closed Universe at so high statistical preference within the $K\Lambda$CDM model, it is virtually impossible to infer constraints consistent with $\Omega_K=0$ from \textit{Planck}+\textit{ext} without paying the price of at least some amount of tension. In the case of \textit{Planck}+\textit{CC} and \textit{Planck}+\textit{Pantheon}, the inference of constraints consistent with $\Omega_K=0$ within $\simeq 1\sigma$ still comes at the cost of a mild internal tension: less than for \textit{Planck}+\textit{FS} and \textit{Planck}+\textit{BAO}, but a tension nonetheless.

While not entirely conclusive, our previous results leave us to entertain the failure of the $K\Lambda$CDM model as the least unlikely implication of our findings. A failure of the $K\Lambda$CDM model does not, however, necessarily imply that the Universe is spatially flat. Rather, it could imply that there might be missing ingredients in the $K\Lambda$CDM model, which need not necessarily have anything to do with spatial curvature. One possibility, considered recently for instance in~\cite{DiValentino:2020hov,DiValentino:2020kpf,Yang:2021hxg,Gonzalez:2021ojp}, is that more freedom in the dark sector might help reconcile \textit{Planck} and external datasets within a non-flat Universe. Examples in this sense include, but are not limited to, considering a dark energy equation of state $w \neq -1$, possibly time-varying, or allowing for interactions between dark matter and dark energy. Therefore, rather than definitely pointing towards the fact that the Universe is spatially flat, the failure of the $K\Lambda$CDM might instead be the sign of richer dynamics in the dark sector, whose composition at present remains unknown.

\section{Conclusions}
\label{sec:conclusions}

The question of what is the shape of the Universe, and more precisely its spatial geometry, is a central one in cosmology, and has been the subject of much debate in recent literature. The apparent preference for a closed Universe from \textit{Planck} CMB temperature and polarization anisotropy power spectra is at odds with a host of complementary precision cosmological data, including BAO and CMB lensing measurements. The debate has centered around both the interpretation of the results obtained from \textit{Planck} data alone~\cite{Efstathiou:2020wem}, and the combination of this dataset with external observations in tension therewith within the assumption of a non-flat Universe~\cite{Handley:2019tkm,DiValentino:2019qzk}. 

In this work, we have instead investigated a different class of cosmological measurements, namely full-shape (FS) galaxy power spectrum measurements from the BOSS DR12 CMASS sample. Combining these measurements with \textit{Planck} CMB data to break the geometrical degeneracy, we have constrained the curvature parameter to be $\Omega_K=0.0023 \pm 0.0028$, which requires the Universe to be spatially flat to sub-percent precision. This finding is in excellent qualitative and quantitative agreement with analogous results obtained combining \textit{Planck} with BAO measurements.

At the same time, as is visually clear from Fig.~\ref{fig:base_omegak_tri}, \textit{Planck} and FS measurements are in tension when assuming a curved Universe. Using the ${\cal I}$ tension diagnostic, based on the deviance information criterion and discussed in Section~\ref{sec:methods}, we find $\log_{10}{\cal I}\approx -2.5$, corresponding to a decisive tension on the Jeffreys-like scale we adopt. A similar level of tension exists between \textit{Planck} and BAO measurements, as already discussed earlier in~\cite{Handley:2019tkm,DiValentino:2019qzk}. This tension suggest that, while FS (and BAO) measurements are important due to their ability to break the geometrical degeneracy once combined with \textit{Planck} data, their combination should be considered with caution within a non-flat Universe.

These results could indicate unaccounted for systematics in one or more of the datasets, or a failure of the simple 7-parameter $\Lambda$CDM+$\Omega_K$ model. In order to discriminate between these two possibilities, we have considered the use of alternative probes to break the geometrical degeneracy, in the form of uncalibrated Hubble flow SNeIa distance moduli, and cosmic chronometer measurements of $H(z)$. The combination of \textit{Planck} with these external datasets is still consistent with spatial flatness, while exhibiting milder internal tensions than those between \textit{Planck} and FS/BAO measurements (partly by virtue of larger error bars). While not entirely conclusive, these results bring us to consider the failure of the $K\Lambda$CDM model as a possible explanation for our results. Note that this does not necessarily imply that the Universe is spatially flat, but might be the sign of missing ingredients in the $\Lambda$CDM+$\Omega_K$ model. Possibilities in this sense include new physics in the dark sector, considered in this context in recent studies~\cite{DiValentino:2020hov,DiValentino:2020kpf,Yang:2021hxg,Gonzalez:2021ojp}.

There is, of course, ample opportunity for following up and improving on our work. We envisage two directions in particular. First of all, it would certainly be worth improving our theoretical modelling of the full-shape galaxy power spectrum beyond our treatment based on \texttt{Halofit} on top of a tree-level model, using for instance 1-loop perturbation theory modelling (as done recently in e.g.~\cite{DAmico:2019fhj,Ivanov:2019pdj,Colas:2019ret,Ivanov:2019hqk,Philcox:2020vvt,DAmico:2020kxu,Nishimichi:2020tvu,Chudaykin:2020aoj,Ivanov:2020ril,DAmico:2020ods,Philcox:2020xbv,Niedermann:2020qbw,Philcox:2020zyp,Chudaykin:2020ghx,Smith:2020rxx}). We expect that on the scales explored, the impact of 1-loop and counterterm corrections to the tree-level power spectrum should be small (see e.g. Fig.~4 in~\cite{Chudaykin:2020aoj}, and recall that our full-shape galaxy power spectrum is measured at an effective redshift of $z_{\rm eff}=0.57$).

Another direction along which it would be interesting to improve our work is to consider extended models. When working within the assumption of a non-flat Universe, it is extremely important to check the stability of one's conclusions against a larger parameter space, as recently remarked in~\cite{DiValentino:2020hov}. The two cosmological parameters most strongly degenerate with $\Omega_K$ are the dark energy equation of state and the sum of the neutrino masses: we plan to check the robustness of our results against extensions of the $K\Lambda$CDM model where these two parameters are allowed to vary in a follow-up work.

Our results open a new window onto the debate concerning the spatial curvature of the Universe and cosmic concordance, and highlight the importance of full-shape galaxy power spectrum measurements in the era of precision cosmology. However, this debate remains open. While it is unlikely that the spatial curvature of the Universe is as large as suggested by \textit{Planck} alone~\cite{Aghanim:2018eyx}, inconsistencies between \textit{Planck} and other datasets, including full-shape galaxy power spectrum and BAO measurements, within the context of a curved Universe, prevent us from asserting with full confidence that the Universe is indeed spatially flat (see e.g.~\cite{Vagnozzi:2020dfn} for further progress in this sense).

\section*{Acknowledgements}
\noindent We thank the anonymous referee for important remarks which helped improve our discussion and interpretation of our results. S.V. thanks George Efstathiou, Steven Gratton, Will Handley, Mikhail Ivanov, and Isabelle Tanseri for very useful discussions. S.V. is supported supported by the Isaac Newton Trust and the Kavli Foundation through a Newton-Kavli Fellowship, and by a grant from the Foundation Blanceflor Boncompagni Ludovisi, n\'{e}e Bildt. S.V. acknowledges a College Research Associateship at Homerton College, University of Cambridge. E.D.V. acknowledges support from an Addison-Wheeler Fellowship awarded by the Institute of Advanced Study at Durham University, and from the European Research Council in the form of a Consolidator Grant with No. 681431. S.G. acknowledges financial support, until September 2020, from the ``Juan de la Cierva-Incorporaci\'on'' program (IJC2018-036458-I) of the Spanish MINECO, from the Spanish grants FPA2017-85216-P (AEI/FEDER, UE), PROMETEO/2018/165 (Generalitat Valenciana), and the Red Consolider MultiDark FPA2017-90566-REDC. Starting from October 2020, he acknowledges support from the European Union's Horizon2020 research and innovation programme under the Marie Skłodowska-Curie grant agreement No. 754496 (H2020-MSCA-COFUND-2016 FELLINI). A.M. is supported by TASP, iniziativa specifica INFN. O.M. is supported by the Spanish grants FPA2017-85985-P, PROMETEO/2019/083 and by the European ITN project HIDDeN (H2020-MSCA-ITN-2019//860881-HIDDeN).

\appendix
\section{Full-shape galaxy power spectrum theoretical modelling and likelihood}
\label{sec:appendix}

In this Appendix, we further discuss our theoretical modelling and likelihood for the full-shape galaxy power spectrum, including the way we account for survey geometry effects. The theoretical modelling and likelihood we describe are the ones some of us developed in earlier works~\cite{Giusarma:2016phn,Vagnozzi:2017ovm,Giusarma:2018jei} as general BOSS full-shape likelihoods. The finer details differ slightly across data releases (we developed it for data releases from DR9 to DR12), and following the general discussion we will briefly discuss the BOSS DR12 case.

For a given set of cosmological parameters, the theoretical value of the full-shape galaxy power spectrum as a function of wavenumber $k$ and measured at an effective redshift $z_{\rm eff}$ (recall for the BOSS DR12 CMASS sample $z_{\rm eff}=0.57$), $P_g^{\rm th}(k,z_{\rm eff})$, is given by:
\small
\begin{eqnarray}
P_g^{\rm th}(k,z_{\rm eff}) = \frac{D_{A,{\rm fid}}^2(z_{\rm eff})}{D_A^2(z_{\rm eff})}\frac{H(z_{\rm eff})}{H_{\rm fid}(z_{\rm eff})} \left ( 1 + \frac{2}{3}\beta + \frac{1}{5}\beta^2 \right )\exp \left [ - \left ( {\hat{k}}\sigma_{\rm FoG} \right ) ^2 \right ]b^2(\hat{k})P_{m,{\rm HF}}(\hat{k},z_{\rm eff})+P_s\,. \nonumber \\
\label{eq:pgth}
\end{eqnarray}
\normalsize
The different terms in Eq.~(\ref{eq:pgth}) account for the Alcock-Paczynski effect, redshift-space distortions, Fingers-of-God, galaxy bias, and a possible incomplete shot noise subtraction. We will now discuss them one by one.

The first two factors on the right-hand side account for the Alcock-Paczynski (AP) effect~\cite{Alcock:1979mp}. This is a geometrical distortion resulting from the fact that, in order to transform the measured redshifts and celestial coordinates (two angles) of a given galaxy catalogue into comoving cartesian coordinates, from which one then estimates the galaxy power spectrum, one needs to assume a reference fiducial cosmology. The AP effect also enters in the rescaled wavenumber $\hat{k}$:
\begin{eqnarray}
\hat{k} = k \left [ \frac{D_A^2(z_{\rm eff})}{D_{A,{\rm fid}}^2(z_{\rm eff})}\frac{H_{\rm fid}(z_{\rm eff})}{H(z_{\rm eff})} \right ] ^{\frac{1}{3}}\,.
\label{eq:hatk}
\end{eqnarray}
Our modelling of the AP effect, which results in small but nonetheless important \%-level corrections, is based on earlier works~\cite{Tegmark:2006az,Reid:2009xm,Parkinson:2012vd} (see especially the Appendix of~\cite{Tegmark:2006az}). The fiducial angular diameter distance and Hubble parameter at the effective redshift of the sample, estimated using the fiducial cosmology assumed by the collaboration, are given by $D_{A\,,{\rm fid}}(z_{\rm eff})$ and $H_{\rm fid}(z_{\rm eff})$ respectively. The factor in round brackets in Eq.~(\ref{eq:pgth}) models linear redshift-space distortions (RSD) due to large-scale peculiar velocities~\cite{Kaiser:1987qv}, usually referred to as the Kaiser effect. In particular, $\beta$ is given by:
\begin{eqnarray}
\beta(\hat{k},z_{\rm eff}) = \frac{f(\hat{k},z_{\rm eff})}{b_0} = \frac{1}{b_0}\frac{d\ln \sqrt{P_m(\hat{k},z_{\rm eff})}}{da}\,,
\label{eq:beta}
\end{eqnarray}
where $P_m$ is the linear matter power spectrum, $b_0$ is the linear galaxy bias parameter (to which we will return later), and $f$ is the logarithmic growth rate, which following~\cite{Lahav:1991wc} we approximate as:
\begin{eqnarray}
f(\hat{k},z_{\rm eff}) \approx \Omega_m(z_{\rm eff})^{0.545} = \frac{H_0^2}{H^2(z_{\rm eff})}\Omega_{m,0}(1+z_{\rm eff})^3 \,,
\label{eq:fkz}
\end{eqnarray}
with $\Omega_{m,0}$ the matter density parameter today. The exponential factor in Eq.~(\ref{eq:pgth}) accounts for the so-called Fingers-of-God (FoG) effect, due to the random motion of virialized objects on small scales~\cite{Jackson:2008yv}, which leads to an exponential suppression in the observed power spectrum below a typical scale related to $\sigma_{\rm FoG}$~\cite{Bull:2014rha}. Since we are considering linear scales, this term has little effect on our analysis. The factor $b(k)$ is the (scale-dependent) galaxy bias, which quantifies the excess clustering of galaxies with respect to the underlying density field. In this work, we consider for simplicity a linear constant galaxy bias model wherein $b(k)=b_0$, which is sufficiently accurate given we will be working on large, linear scales~\cite{Desjacques:2016bnm}. In principle, this prescription can be improved by going beyond the simple constant bias model, as done in a number of recent works such as~\cite{More:2014uva,Amendola:2015pha,Beutler:2016zat,Simon:2017osp,Giusarma:2018jei}. Furthermore, $P_{m\,,{\rm HF}}(\hat{k},z_{\rm eff})$ is the mildly non-linear matter power spectrum computed using the Boltzmann solver \texttt{CAMB} and the \texttt{Halofit} prescription~\cite{Smith:2002dz}.~\footnote{When the neutrino mass is allowed to vary, the matter power spectrum $P_m$ should be replaced by the cold DM plus baryons power spectrum $P_{cb}$, as advocated by a number of works~\cite{Castorina:2013wga,Raccanelli:2017kht,Munoz:2018ajr,Vagnozzi:2018pwo,Valcin:2019fxe,Xu:2020fyg,DePorzio:2020wcz}.} Finally, the term $P_s$ accounts for a potential insufficient shot noise subtraction when computing the galaxy power spectrum starting from the galaxy catalogue. We refer to this term simply as shot noise.

So far we have discussed our modelling of the theoretical galaxy power spectrum $P_g^{\rm th}$ given by Eq.~(\ref{eq:pgth}). However, the resulting power spectrum is not yet ready to be confronted with the measured power spectrum in the \textit{FS} likelihood: one needs to account for the fact that the finite survey geometry introduces mode-coupling between different $k$ modes, which would otherwise be independent. This means that the measured power spectrum is not the true underlying galaxy power spectrum given by Eq.~(\ref{eq:pgth}), but is given by a convolution between the latter and the so-called window function. In practice, since the galaxy power spectrum is measured at a discrete set $k$-bands $k_i$, the window function is more precisely a window matrix $W_{ij} = W(k_i,k_j)$, where the off-diagonal terms capture the mode-coupling between different $k$ modes. The effect of the survey geometry can be understood by studying a (simulated) unclustered random catalog matching the observed survey geometry. We denote the power spectrum of this unclustered random catalog, which we refer to as the ``window power spectrum'' (this is basically the power spectrum of the survey mask), by $P_w(k)$. Finally, the finite survey geometry also leads to an underestimation of the power in modes whose wavelength approaches the size of the survey, an effect which is usually corrected by the so-called ``integral constraint'' (in real space) or ``window subtraction'' (in Fourier space), to be discussed shortly~\cite{Percival:2006gs,deMattia:2019vdg}.

Once survey geometry effects are accounted for, we obtain the convolved theoretical galaxy power spectrum measured at a discrete set $k$-bands $k_i$, $P_g^{\rm conv}(k_i)$. We model this quantity, which is almost ready to be compared to the measured power spectrum in the \textit{FS} likelihood, as follows:
\begin{eqnarray}
P_g^{\rm conv}(k_i) = \sum_{ij}W_{ij}P_g^{\rm th}(k_j) - \frac{\sum_jW_{0j}P_g^{\rm th}(k_j)}{P_w(0)}P_w(k_i)\,,
\label{eq:pgconv}
\end{eqnarray}
where we refer to $W_{0j} = W(0,k_j)$ as the $0$-window function, which accounts for mode-coupling between the $k=0$ and $k_j$ modes, whereas $P_w$ is the window power spectrum discussed previously, with $P_w(0)$ the same quantity measured at $k=0$. The second term on the left-hand side of Eq.~(\ref{eq:pgconv}) is referred to as the ``window subtraction'', and is closely related to the so-called ``integral constraint'' relevant for the real-space 2-point correlation function~\cite{Percival:2006gs,deMattia:2019vdg}. The integral constraint arises because, when computing the galaxy power spectrum, one estimates the average galaxy density from the sample itself. This amounts to the assumption that the mean galaxy density of the survey is equal to the mean galaxy density of the Universe, or in other words that the integral of the inferred density fluctuations across the survey geometry is zero. The non-zero sample variance expected at wavelengths which approach the size of the survey, however, tells us that this should introduce a bias in the measured power spectrum, which is what the integral constraint is modelling. In Fourier space, estimating the average galaxy density from the sample itself means that one is artificially setting $P_g(k=0)=0$, and the window function will propagate this incorrect estimation to modes relevant for cosmological measurements. More precisely, the galaxy power spectrum one is measuring is not the true galaxy power spectrum, but a galaxy power spectrum with the property $P_g \to 0$ as $k \to 0$~\cite{1991MNRAS.253..307P}. We account for the integral constraint in Eq.~(\ref{eq:pgconv}) following earlier works~\cite{Ross:2012sx,Beutler:2013yhm} (see e.g.\ Section~3.3 in~\cite{Ross:2012sx}). We can see that $P_g^{\rm conv}$ as defined in Eq.~(\ref{eq:pgconv}) satifies $P_g^{\rm conv}(k_i=0)=0$ by construction, and thus accounts for the integral constraint bias.

So far we discussed the theoretical modelling of the galaxy power spectrum, which led to $P_g^{\rm th}$ in Eq.~(\ref{eq:pgth}), which we then convolved with the survey window function and corrected for the integral constraint, leading to $P_g^{\rm conv}$ in Eq.~(\ref{eq:pgconv}). The final manipulation required before we can compute the \textit{FS} likelihood is to model the effect of systematics on the galaxy power spectrum. Systematics affecting the BOSS full-shape galaxy power spectrum measurements were studied in detail in~\cite{Ross:2012qm}, where it was found that the strongest systematic impacting the BOSS galaxy density field is related to the local stellar density.

\begin{figure}
\centering
\includegraphics[width=0.7\linewidth]{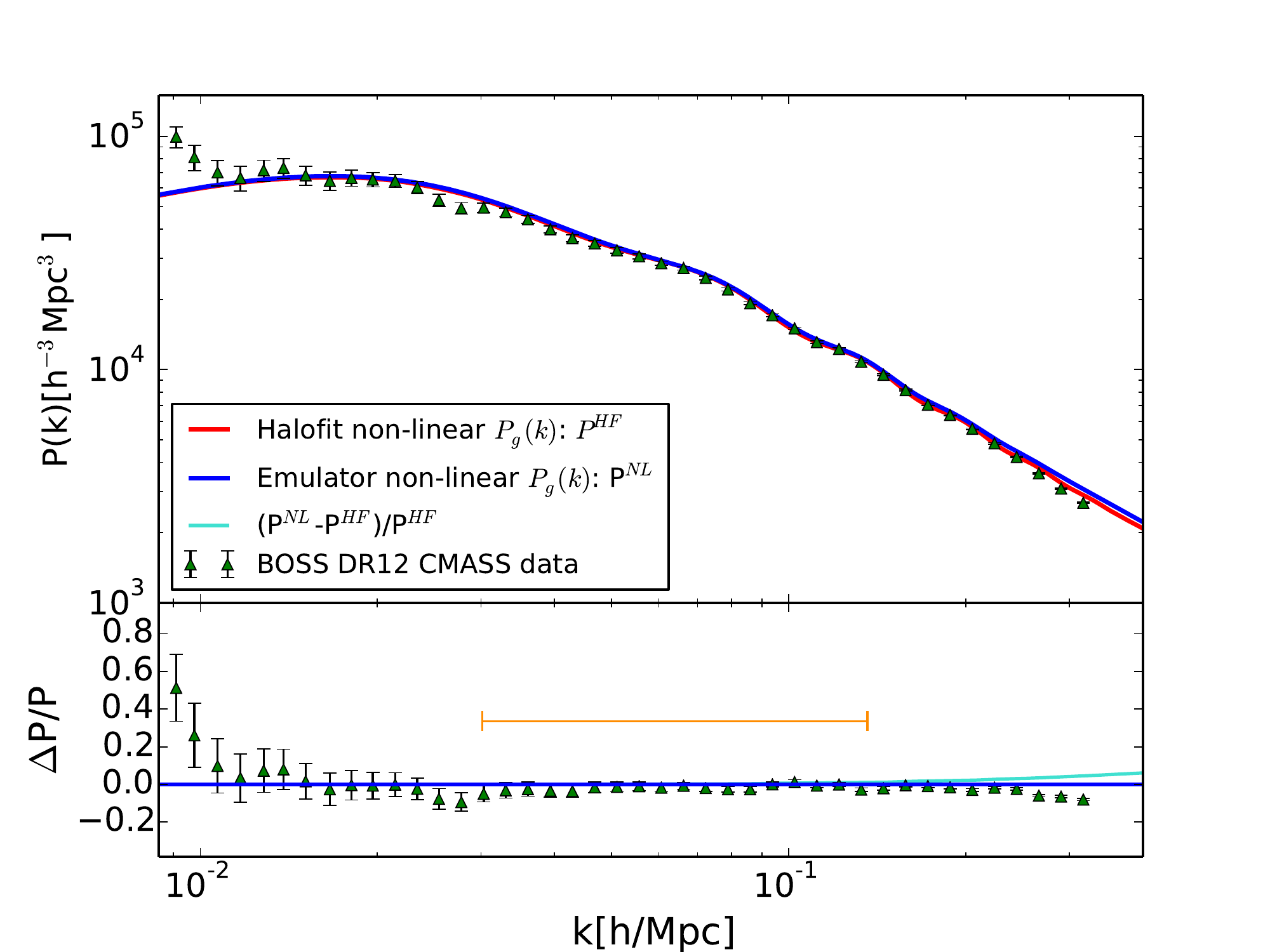}
\caption{\textit{Top panel}: non-linear galaxy power spectrum computed using either the \texttt{Halofit} prescription on top of the linear power spectrum calculated using \texttt{CAMB} (blue curve, $P^{HF}$), or the \textit{Coyote} emulator (red curve, $P^{NL}$), assuming the best-fit cosmological and nuisance parameters from a fit of the $\Lambda$CDM model to \textit{Planck}+\textit{FS}, with the \textit{FS} measurements given by the green datapoints. \textit{Bottom panel}: relative difference between $P^{HF}$ and $P^{NL}$ (turquoise curve), with green datapoints denoting the residuals of the \textit{FS} measurements with respect to $P^{HF}$, for the same choice of cosmological and nuisance parameters. The orange horizontal line denotes the $k$ range $0.03\,h\,{\rm Mpc}^{-1}<k<0.135\,h\,{\rm Mpc}^{-1}$ used in our analysis.}
\label{fig:plot_pk_residuals_coyote_emulator}
\end{figure}

Given that this systematic, and more generally other systematics due for instance to fiber collisions and missing close-pairs, are relatively well understood by the BOSS collaboration~\cite{Ross:2016gvb}, these are modelled as systematic weights applied to each galaxy (see e.g.\ Eq.~(18) in~\cite{Anderson:2013zyy}), which multiply the usual Feldman-Kaiser-Peacock (FKP) weights used to compute the galaxy power spectrum following the widely used FKP prescription first developed in~\cite{Feldman:1993ky}. The measured galaxy power spectrum, $P_g^{\rm meas}(k)$, is computed with all weights applied. However, using the same pipeline one can also compute a ``no-systematics'' power spectrum, $P_g^{\rm nosys}(k)$, which is obtained by only applying FKP weights and not the systematics ones. Following~\cite{Ross:2012sx} we assume that the correction for systematics always has the same form, given by $P_g^{\rm meas}(k)-P_g^{\rm nosys}(k)$, but its amplitude can vary. In other words, we fix the form of the systematic correction in the observed power spectrum, but our model is flexible enough to account for the possibility that the measurement has either oversubtracted the systematic bias or that there remains a residual systematic bias, by rescaling the amplitude thereof.

Following~\cite{Ross:2012sx}, we model the systematics-corrected convolved theoretical galaxy power spectrum, $P_g^{\rm sys}$, as follows:
\begin{eqnarray}
P_g^{\rm sys}(k) = P_g^{\rm conv}(k) + S \left [ P_g^{\rm meas}(k)-P_g^{\rm nosys}(k) \right ]\,,
\label{eq:pgsys}
\end{eqnarray}
where $S$ is a nuisance parameter describing the amount of systematic correction required by the data. In particular, $S=0$ represents the fiducial case where any systematic bias has been correctly removed from the measurement, whereas $S>0$ account for the possibility that a systematic bias has been incorrectly oversubtracted, and finally $S<0$ corresponds to the case where a systematic bias remains.

The quantity $P_g^{\rm sys}$ in Eq.~(\ref{eq:pgsys}) is ready to be compared against the measured galaxy power spectrum $P_g^{\rm meas}$ in the \textit{FS} likelihood. Specifically, the \textit{FS} log-likelihood, denoted by $\ln {\cal L}_{FS}$, is given by the following expression:
\begin{eqnarray}
\ln {\cal L}_{FS} = -\frac{\Delta^TC^{-1}\Delta}{2}\,, \quad \Delta \equiv P_g^{\rm meas}-P_g^{\rm sys}\,,
\label{eq:fsloglikelihood}
\end{eqnarray}
where $C$ is the covariance matrix for the BOSS DR12 CMASS galaxy power spectrum measurements, estimated using mocks of the sample generated from the quick particle mesh algorithm~\cite{White:2013psd}.

In summary, given a set of cosmological parameters, the steps required in going from the matter power spectrum $P_{m,{\rm HF}}$ computed using \texttt{CAMB} to the \textit{FS} likelihood are given by the following:
\begin{eqnarray}
&&P_{m,{\rm HF}}\,\, [{\texttt{CAMB}}] \to P_g^{\rm th}\,\, [\text{Eq.~(\ref{eq:pgth})}] \to P_g^{\rm conv}\,\, [\text{Eq.~(\ref{eq:pgconv})}] \nonumber \\
\to &&P_g^{\rm sys}\,\, [\text{Eq.~(\ref{eq:pgsys})}] \to \Delta\,\, [\text{Eq.~(\ref{eq:fsloglikelihood})}] \to \ln {\cal L}_{FS}\,\, [\text{Eq.~(\ref{eq:fsloglikelihood})}]\,.
\label{eq:steps}
\end{eqnarray}
The minimal implementation of our \textit{FS} likelihood therefore features 4 nuisance parameters: the linear bias $b_0$ [Eq.~(\ref{eq:pgth})], the FoG parameter $\sigma_{\rm FoG}$ [Eq.~(\ref{eq:pgth})], the shot noise term $P_s$ [Eq.~(\ref{eq:pgth})], and the systematics amplitude $S$ [Eq.~(\ref{eq:pgth})]. We adopt flat priors on all these parameters, specifically within the ranges $b_0 \in [0,5]$, $\sigma_{\rm FoG} \in [4,10]\,{\rm Mpc}$, $P_s \in [0,10000]\,h{-3}\,{\rm Mpc}^3$, and $S \in [-1,1]$. We analyze the modes within the range $0.03\,h\,{\rm Mpc}^{-1}<k<0.135\,h\,{\rm Mpc}^{-1}$. The choice of large-scale cut $k=0.03\,h\,{\rm Mpc}^{-1}$ is driven by the observation that large-scale clustering of the BOSS galaxies is affected by the earlier described systematics (in particular stellar density). On the other hand, the choice of small-scale cut $k=0.135\,h\,{\rm Mpc}^{-1}$ is made to reduce the impact of non-linearities, so that the \texttt{Halofit} prescription is reliable. Recall that modes at $z=0$ start to become mildly non-linear at $k=0.12\,h\,{\rm Mpc}^{-1}$, but the fact that the BOSS DR12 CMASS sample is at a higher effective redshift $z_{\rm eff}=0.57$, where any given mode is less in the non-linear regime than at $z=0$, allows us to push to slightly smaller scales. For a more complete analysis of the impact of systematics and non-linearities on our modelling, we refer the reader to~\cite{Giusarma:2016phn,Vagnozzi:2017ovm}.

Our model for the FS monopole presented in this Appendix was validated against an emulator for the fully non-linear galaxy power spectrum in~\cite{Giusarma:2016phn,Vagnozzi:2017ovm}. We show this in Fig.~\ref{fig:plot_pk_residuals_coyote_emulator}, where we compare the BOSS DR12 CMASS FS measurements against our theoretical model based on \texttt{Halofit}, and the predictions for the non-linear galaxy power spectrum from the \textit{Coyote} emulator~\cite{Heitmann:2008eq,Heitmann:2013bra,Kwan:2013jva}, both computed assuming the best-fit cosmological and nuisance parameters from a fit of the $\Lambda$CDM model to \textit{Planck}+\textit{FS} dataset combination. We run the \textit{Coyote} emulator adopting the halo occupation distribution model for SDSS LRGs, best suited for the BOSS DR12 CMASS sample. From Fig.~\ref{fig:plot_pk_residuals_coyote_emulator} wee see that the adopted $k$ range $0.03\,h\,{\rm Mpc}^{-1}<k<0.135\,h\,{\rm Mpc}^{-1}$ is safe from both observational systematics on large scales and large non-linear corrections on small scales. Note that, compared to our earlier works~\cite{Giusarma:2016phn,Vagnozzi:2017ovm,Giusarma:2018jei}, we have chosen a significantly more conservative $k_{\max}$ cutoff.

Earlier, we mentioned that this model is the one some of us developed in earlier works~\cite{Giusarma:2016phn,Vagnozzi:2017ovm,Giusarma:2018jei} as general BOSS full-shape likelihoods, with the finer details varying across data releases. The complete likelihood we described is the one we adopted for our independent DR9 analyses. On the other hand, for the DR12 analyses the integral constraint files were not publicly available, and therefore the window subtraction term in Eq.~(\ref{eq:pgconv}) is not applied. The reason, as can be seen in Appendix~A2 of~\cite{Beutler:2016arn}, is that the integral constraint correction in BOSS only affects modes $k \lesssim 0.005\,h\,{\rm Mpc}^{-1}$, which are beyond the scale cuts we apply. In addition, the ``no-systematics'' power spectrum was not available for DR12.

\bibliography{Curvature_full_shape}
\bibliographystyle{JHEP}

\end{document}